\documentclass[12pt,preprint]{aastex}
\usepackage{rotating}
\usepackage{epsfig}
\usepackage{ifpdf}

\begin{document}

\title{The Radio Jet Associated with the Multiple V380 Ori System}

\author{
Luis F. Rodr\'\i guez\altaffilmark{1}, J. Omar Yam\altaffilmark{1,2}, Carlos Carrasco-Gonz\'alez\altaffilmark{1},  Guillem Anglada\altaffilmark{3}
and Alfonso Trejo\altaffilmark{4}}

\altaffiltext{1}{Instituto de Radioastronom\'\i a y Astrof\'\i sica, 
UNAM, Apdo. Postal 3-72 (Xangari), 58089 Morelia, Michoac\'an, M\'exico}
\altaffiltext{2}{Departamento de Ciencias, Universidad de Quintana Roo, Boulevard Bah\'\i a s/n, Col. Del Bosque, 77019 Chetumal, Quintana Roo,
M\'exico} 
\altaffiltext{3}{Instituto de Astrof\'\i sica de Andaluc\'\i a, CSIC, Glorieta de la Astronom\'\i a, s/n, E-18008, Granada, Spain}
\altaffiltext{4}{Academia Sinica Institute of Astronomy and Astrophysics, P.O. Box 23-141, Taipei 10617, Taiwan}
\email{
l.rodriguez@crya.unam.mx}
 
\begin{abstract}
The giant Herbig-Haro object 222 extends over $\sim$6$'$ in the plane of the sky, with a bow shock morphology.
The identification of its exciting source has remained uncertain over the years. A non-thermal radio source 
located at the core of the shock structure was proposed to 
be the exciting source. However, Very Large Array studies showed that
the radio source has a clear morphology
of radio galaxy and a lack of flux variations or proper motions, favoring an extragalactic origin. 
Recently, an optical-IR study proposed that this giant HH object is driven by the multiple stellar system V380 Ori,
located about 23$'$ to the SE of HH 222. The exciting sources of HH systems are usually detected
as weak free-free emitters at centimeter wavelengths. Here we report the detection of an elongated
radio source associated with the Herbig Be star or with its close infrared companion in the multiple V380 Ori system.  This radio source has the characteristics of a thermal radio jet and is aligned with the
direction of the giant outflow defined by HH~222 and its suggested counterpart to the SE, HH~1041. We propose that this radio jet traces the origin of the large scale HH outflow. 
Assuming that the jet
arises from the Herbig Be star, the radio luminosity is a few times smaller than the value expected from the radio-bolometric correlation for
radio jets, confirming that this is a more evolved object than those used to establish the correlation.

\bigskip
{\bf ~~~~~~~~~~~~~~~To appear in The Astronomical Journal}

\end{abstract}  


\section{Introduction}

The very large HH object 222 (also known as the Orion streamers) is located in the northern part of the
L1641 dark cloud in Orion (Cohen \& Schwartz 1983), near the region where the classic systems HH 1/2 and 
HH 34 are found. Over the years, different sources have been proposed to excite this shocked region.
Reipurth \& Sandell (1985) proposed that a wind from the T Tau star V571 Ori, at only 15$''$ to the NW of
the core of HH 222,
was impacting the edge of a cloud and producing the HH object. A few years later, Yusef-Zadeh et al. (1990)  
and Morgan et al. (1990) independently detected a strong non-thermal radio source at the core of HH 222 
that was proposed as its true exciting source.

Other arguments, however, gravitated against the latter interpretation. Reipurth et al. (1993) did not detect a
1300 $\mu$m source in association with the non-thermal radio source. The embedded exciting sources of
HH systems typically exhibit detectable radiation at these millimeter wavelengths. 
Castets et al. (2004) mapped the region around HH 222 in
several molecular transitions at mm wavelengths, but failed to detect any dense
core that could contain a young driving source.
Finally, Trejo \& Rodr\'\i guez (2010) obtained new high-resolution
6 cm and 20 cm continuum images of the radio source
and found it exhibits the double-lobe morphology characteristic of radio galaxies and
that comparing data taken with a separation of 17 years there was no evidence of 
changes in the flux density and morphology or detectable proper motions 
at the levels expected for a jet source at the Orion distance.
These authors concluded that the non-thermal radio
source most probably is a radio galaxy aligned by chance with the line of sight to
HH 222. The search for the exciting source of HH 222 was on again.

In Figure 1 we show the 4.86 GHz VLA data for the HH~222 region discussed by Trejo \& Rodr\'\i guez (2010).
In this image we have concatenated the 1991 and 2008 data and self-calibrated it in phase and amplitude.
It is clearly seen from its morphology that the double non-thermal radio source is almost certainly a radio galaxy
whose central infrared counterpart is the source IRS 1, first noted by Yusef-Zadeh et al. (1990) and discussed in more detail
by Reipurth et al. (2013). As proposed by the latter authors, IRS 1 is quite likely an elliptical galaxy powering the radio galaxy.
The nearby source IRS 2 (see Figure 1), also reported by  Yusef-Zadeh et al. (1990) and
Reipurth et al. (2013), has no associated radio emission at a 3-$\sigma$ upper limit of 69 $\mu$Jy.

\begin{figure}
\centering
\vspace{-3.8cm}
\includegraphics[angle=0,scale=0.8]{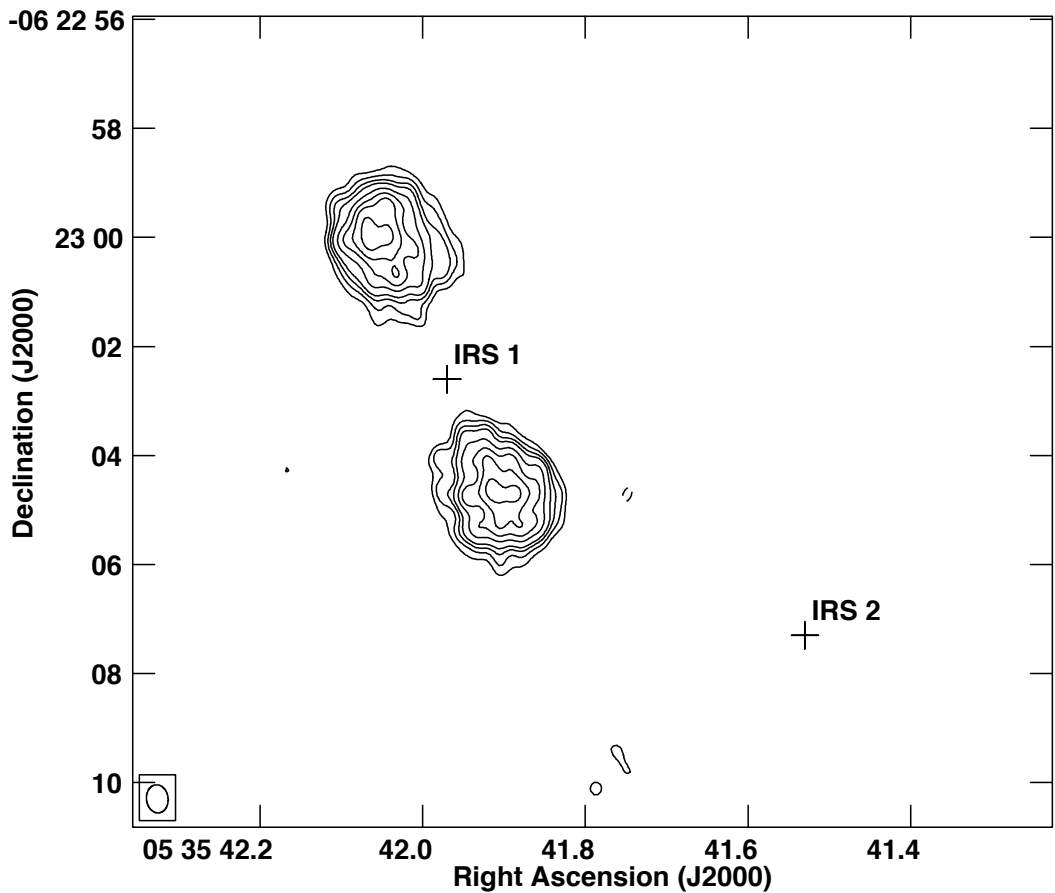}
\caption{\small VLA 4.86 GHz  continuum contour image of the core of the HH~222 region, made from the 1991 and 2008 data discussed by
Trejo \& Rodr\'\i guez (2010). 
The contours are -4, -3, 3, 4, 5, 6, 8, 10, 12 and 15 times 23
$\mu$Jy beam$^{-1}$, the rms noise of the
image. The classic double lobe morphology of a radio galaxy is evident.
The half-power contour of the synthesized beam  is shown in the bottom left corner
($0\rlap.{''}52 \times 0\rlap.{''}40;  PA = +7^\circ$). 
The crosses mark the positions of the two infrared sources discussed by Yusef-Zadeh et al. (1990) and Reipurth et al. (2013).
}
\label{fig1}
\end{figure}


Reipurth et al. (2013) did a detailed optical and near-IR continuum and line
study of HH 222 and found that it has proper motions of 107$\pm$25 km s$^{-1}$ toward
P.A. = 329$^\circ$$\pm$10$^\circ$. These results pointed to a source to the SE of HH 222 as
its exciting source. These authors proposed that the most likely source was 
V380 Ori (= BD-06$^\circ$1253 = Parenago 2393 = HBC 164 = Haro 4-235), located at $22\rlap.{'}6$ to the SE
of HH 222. V380 Ori is a hierarchical quadruple system composed of a Herbig Be star with an infrared companion 
separated by $0\rlap.{''}15$ toward a P.A. of $204^\circ$ that was discovered by Leinert et al. (1997).
These two components were labeled respectively as Aa and Ab by Reipurth et al (2013). Component Aa is itself a spectroscopic binary
with the Herbig Be star having
an effective temperature of 10500$\pm$500 K and a luminosity of
$\sim$200 $L_\odot$ and the spectroscopic companion having an effective temperature of 5500$\pm$500 K and a luminosity of
$\sim$3 $L_\odot$ (Alecian et al. 2009). 
Here we will consider the spectroscopic binary as a single object.
A fourth component
of spectral type M5 or M6 (labeled B by Reipurth et al. 2013) is located at a distance of $8\rlap.{''}8$ at a P.A. of $120^\circ$.

Reipurth et al. (2013) also found a new Herbig-Haro object, HH 1041, located at $17\rlap.{'}3$ to the SE of V380 Ori 
in the opposite direction of HH 222 and likely
forming part of a counterflow to HH 222. Figure 14 of Reipurth et al (2013) shows the positions of HH 222, V380 Ori and HH 1041.

Since the exciting sources of HH objects are known to be practically always associated with
faint free-free emission at centimeter wavelengths (e.g., Rodr\'\i guez \& Reipurth 1998), we did a search for such a source concatenating
several epochs of observation from the VLA
archives to achieve the highest sensitivity possible and we also obtained new observations using the ultrasensitive Jansky VLA. 

\section{Observations}

The VLA archive observations were made at C-band (4.86 GHz) during 11 epochs that are summarized in Table 1.
The average epoch of these data is 1998.01.
These observations were made with the phase center at or very close the position of HH~1/2 VLA~1 ($\alpha(J2000) = 05^h~ 36^m~ 22\rlap.^s84$;
$\delta(J2000)$ = $-$06$^\circ~ 46'~ 06\rlap.{''}2$), the exciting source of the
HH~1/2 system (Pravdo et al. 1985; Rodr\'\i guez et al. 2000).
The data were calibrated following the standard procedures in the AIPS  (Astronomical Image Processing System)
software package of NRAO\footnote{The National 
Radio Astronomy Observatory is a facility of the National Science Foundation operated
under cooperative agreement by Associated Universities, Inc.}  and concatenated in a single file. An image with weighting of ROBUST = 5 (equivalent to natural
weighting in AIPS; Briggs 1995) was made to optimize the sensitivity of the image, at the expense of losing some 
angular resolution.
A source was clearly detected in close vicinity ($\sim 0\rlap.{''}1$) of the Herbig Be star V380 Ori (see Figure 2). Since the observations were made with 
bandwidths of 50 MHz (two of them, adjacent in frequency) and V380 Ori is located about $3\rlap.'2$ to the NE of HH~1-2 VLA1,
the radio source presents significant bandwidth smearing. Under these limitations, the source appears
to be spatially unresolved ($\leq2\rlap.{''}0$). Its total flux density is 0.20$\pm$0.03 mJy. All flux densities presented here
have been corrected for the primary beam response.

\begin{figure}
\centering
\vspace{-3.8cm}
\includegraphics[angle=0,scale=0.8]{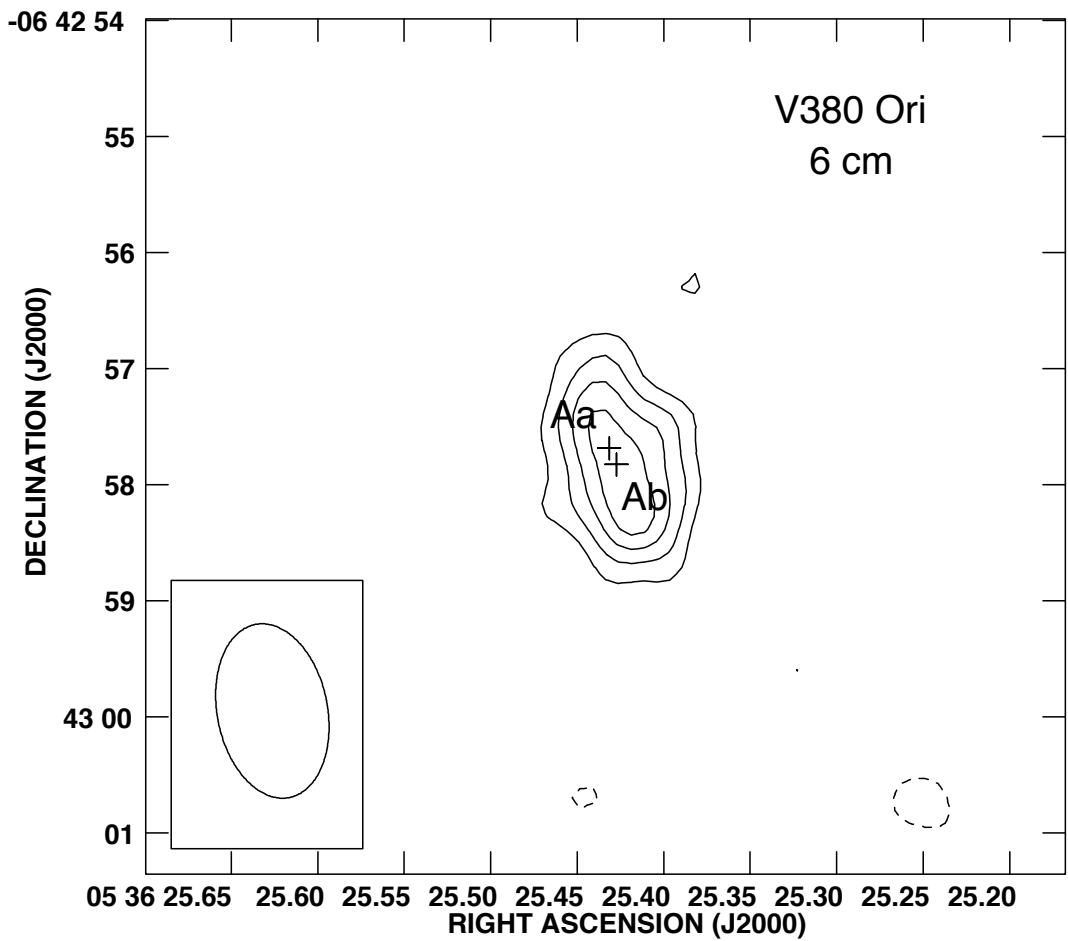}
\vskip -3.0cm
\caption{\small VLA 4.86 GHz  continuum contour image of the V380 Ori region, made from the archive data summarized in Table 1.
The contours are -4, -3, 3, 4, 5, and 6 times 30.7
$\mu$Jy beam$^{-1}$. The image has been corrected for the response of the primary beam. 
The half-power contour of the synthesized beam has been corrected to include the
effect of bandwidth smearing and is shown in the bottom left corner
($1\rlap.{''}52 \times 0\rlap.{''}95;  PA = +11^\circ$).
In this image the source appears unresolved and with a total flux density of
0.20$\pm$0.03 mJy. The positional accuracy of the VLA image is $0\rlap.{''}3$. The crosses mark the position of the Herbig Be star 
(component Aa) from
van Leeuwen (2007) and from the infrared companion (component Ab) as
derived from the offsets of Leinert et al. (1997).}
\label{fig2}
\end{figure}


The new observations were made with the Karl G. Jansky Very Large Array of NRAO in the C (4.4 to 6.4 GHz) and X (7.9 to 9.9 GHz) bands
during 2012 May 26, under project 12A-240.
At that time the array was in its B configuration.  The phase center was 
at $\alpha(2000) = 05^h~ 36^m~ 22\rlap.^s00$;
$\delta(2000)$ = $-$06$^\circ~ 46'~ 07\rlap.{''}0$. The absolute amplitude calibrator was 0137+331 and
the phase calibrator was J0541$-$0541.

The digital correlator of the JVLA was configured at each band in 16 spectral windows of 128 MHz width each subdivided 
in 64 channels of 2 MHz. The narrow width of each channel eliminates the bandwidth smearing that
limited observations away from the phase center in the classic VLA.
The total bandwidth of the observations was about 2.048 GHz in a full-polarization mode.
The data were analyzed in the standard manner using the CASA (Common Astronomy Software Applications) package of NRAO,
although for some stages of the analysis we used the AIPS 
package. 
For all the imaging, we used the ROBUST parameter of CLEAN set to 2 (equivalent to natural weighting in CASA), to obtain a better sensitivity.
 
We confirm the detection of the radio source  in close vicinity to V380 Ori. Its total flux densities were 0.14$\pm$0.03 mJy and 0.25$\pm$0.04 mJy, for the
C and X bands, respectively. Assuming a power law for the emission, the radio flux density of V380 Ori for epoch 2012 May 26 can be described as

$$\Biggl[{{S_\nu} \over {{\rm mJy}}}\Biggr] = 0.14\pm0.03 \Biggl[ {{\nu} \over {5.4~{\rm GHz}}}\Biggr]^{1.2\pm0.5}.$$

An image of the radio source was made combining the C and X band data, covering an almost continuous 
frequency range from 4.4 to 9.9 GHz. The image is shown in Figure 3.
We searched unsuccessfully for additional sources in the region. In particular, we did not detect radio emission from component B at a 3-$\sigma$ level of 34 $\mu$Jy.

\begin{figure}
\centering
\vspace{-2.8cm}
\includegraphics[angle=0,scale=0.70]{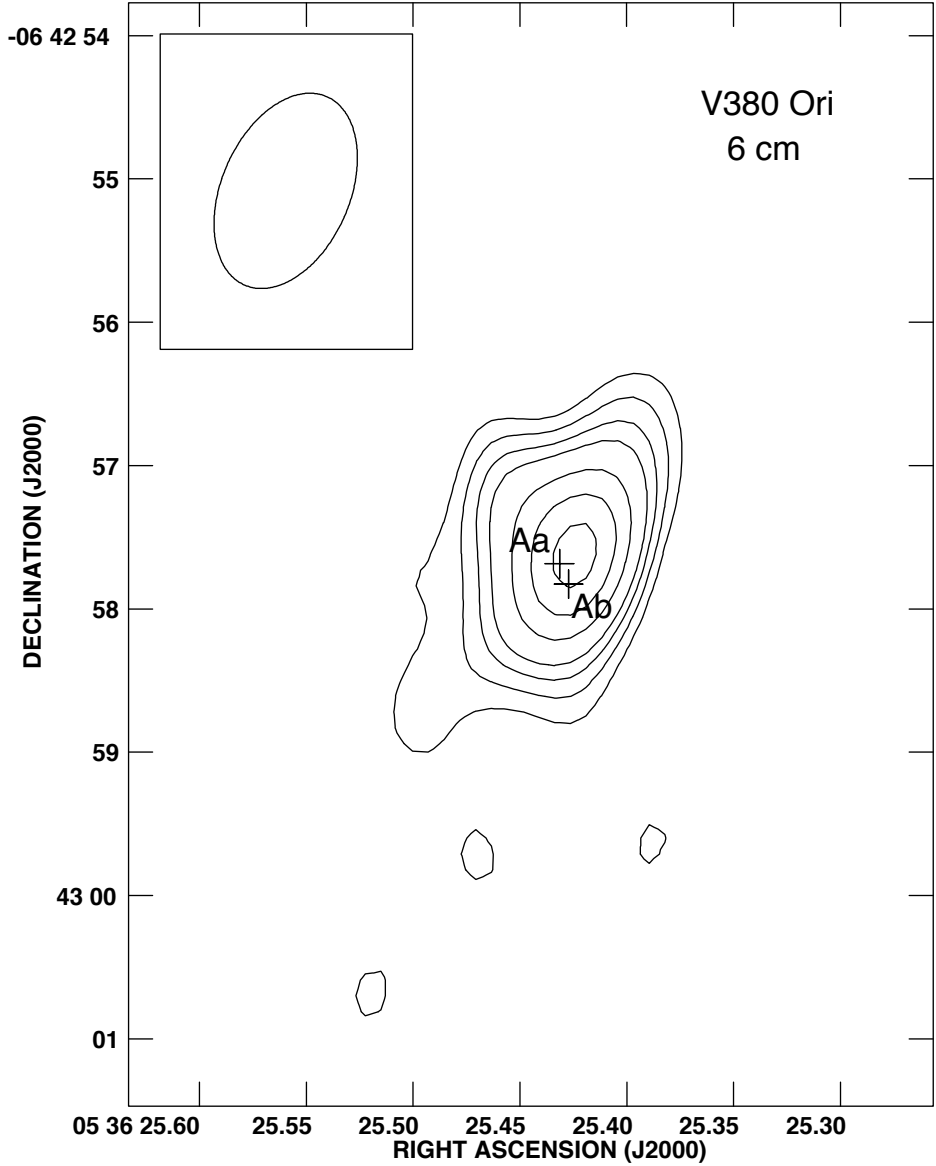}
\vskip-1.5cm
\caption{\small JVLA 6 cm  continuum contour image of the V380 Ori region, made from the new (2012) Jansky VLA observations.
The contours are -4, -3, 3, 4, 5, 6, 8, 10 and 12 times 11.2
$\mu$Jy beam$^{-1}$, the rms noise of this region of the
image. The image has been corrected for the response of the primary beam. 
The half-power contour of the synthesized beam  is shown in the bottom left corner
($1\rlap.{''}43 \times 0\rlap.{''}89;  PA = -24^\circ$).
In this image the source appears marginally resolved with deconvolved dimensions  of
$1\rlap.{''}3 \pm 0\rlap.{''}4 \times 0\rlap.{''}8 \pm 0\rlap.{''}3; PA = +161^\circ \pm 36^\circ$ and with a total flux density of
0.22$\pm$0.03 mJy. The positional accuracy of the VLA image is $0\rlap.{''}1$. The crosses mark the position of the Herbig Be star 
(component Aa) from
van Leeuwen (2007) and from the infrared companion (component Ab) as
derived from the offsets of Leinert et al. (1997).}
\label{fig3}
\end{figure}

\pagebreak

\begin{deluxetable}{c c c c c c c}
\tabletypesize{\scriptsize}
\tablecaption{Parameters of VLA archive observations in C-band}
\tablehead{                        
\colhead{Epoch}                              &
\colhead{   }                              &
\colhead{   }                              &
\multicolumn{2}{c}{Phase Center} &
\colhead{Amplitude}                     &   
\colhead{Phase}                     \\
\colhead{(yy/mm/dd)}                              &
\colhead{Project}                              &
\colhead{Configuration}                              &
\colhead{$\alpha$(J2000)}    &
\colhead{$\delta$(J2000)}    &
\colhead{Calibrator}                     &
\colhead{Calibrator}                     
}
\startdata

84/10/02 & AR114 & D &  $05^h~36^m~23\rlap.^s865$ & $-06^\circ~45{'}~11\rlap.{''}66$ & 3C286  &  0539-057 \\
85/11/21 & AR131 & D & $05^h~36^m~22\rlap.^s846$ & $-06^\circ~46{'}~08\rlap.{''}59$ & 3C286  &  0539-057 \\
86/01/13 & AR131 & D &  $05^h~36^m~22\rlap.^s846$ & $-06^\circ~46{'}~08\rlap.{''}59$ & 3C286  &  0539-057 \\
86/03/01 & AR131 & A &  $05^h~36^m~22\rlap.^s846$ & $-06^\circ~46{'}~08\rlap.{''}59$ & 3C286  &  0539-057 \\
86/03/03 &  AR131 & A & $05^h~36^m~22\rlap.^s846$ & $-06^\circ~46{'}~08\rlap.{''}59$ & 3C286  &  0539-057 \\
86/09/12 &  AR131 & A & $05^h~36^m~22\rlap.^s846$ & $-06^\circ~46{'}~08\rlap.{''}59$ & 3C286  &  0539-057 \\
86/09/13 &  AR131 & A & $05^h~36^m~22\rlap.^s846$ & $-06^\circ~46{'}~08\rlap.{''}59$ & 3C286  &  0539-057 \\
86/12/09 &  AR131 & C & $05^h~36^m~22\rlap.^s846$ & $-06^\circ~46{'}~08\rlap.{''}59$ & 3C286  &  0539-057 \\
92/11/02 & AR278 & A  & $05^h~36^m~22\rlap.^s846$ & $-06^\circ~46{'}~08\rlap.{''}59$ & 3C286  &  0539-057 \\
92/12/18 & AR278 & A  & $05^h~36^m~22\rlap.^s846$ & $-06^\circ~46{'}~08\rlap.{''}59$ & 3C286  &  0539-057 \\
92/12/19 & AR278 & A  & $05^h~36^m~22\rlap.^s846$ & $-06^\circ~46{'}~08\rlap.{''}59$ & 3C286  &  0539-057 \\

\enddata
\end{deluxetable}

\pagebreak

\section{The nature of the radio source detected in close vicinity to V380 Ori}

\subsection{A radio jet}

Several characteristics of the radio source favor an interpretation in terms of a thermal radio jet (Anglada 1996; Rodr\'\i guez 1997; Eisloffel et al. 2000; Anglada et al. 2015):

i) The spectral index of 1.2$\pm$0.5 is consistent with the partially optically thick free-free emission produced by thermal jets.

ii) The source exhibits little or no temporal flux density variation, as generally observed in this type of sources. Most of the radio jets that have been monitored 
over the years show no evidence of variability
above the 10-20\% level (e.g. Rodr\'\i guez et al. 2008; Loinard et al. 2010; Carrasco-Gonz\'alez et al. 2012;
Rodr\'\i guez et al. 2014).

iii) The source in the combined C and X band image has deconvolved dimensions of $1\rlap.{''}3 \pm 0\rlap.{''}4 \times 0\rlap.{''}8 \pm 0\rlap.{''}3; PA = +161^\circ \pm 36^\circ$,
with the position angle consistent modulo $180^\circ$ with the position angle of the proper motions of HH~222 ($329^\circ \pm 10^\circ$; Reipurth et al.
2013) and the direction from V380 Ori to the core
of HH 222 ($\simeq331^\circ$). We propose that this radio source traces the origin of the HH system.

\subsection{Is the radio source associated with the Herbig Be star in V380 Ori or with its infrared companion?}

Our radio images show that the peak of the radio emission falls within $\sim 0\rlap.{''}1$ 
of the optical position of the Herbig Ae star as given by van Leeuwen (2007). 
Unfortunately, the quality of the available astrometry is insufficient to establish if
the radio jet is associated with the Herbig Be star (Aa) or with its close infrared companion (Ab). The primary star (Aa) hosts a dipole
magnetic field with polar strength of $\sim$2 kG (Alecian et al. 2009) suggesting the possibility of gyrosynchrotron emission from this star.
However, the characteristics of the observed radio emission favor a free-free nature.
Leinert et al. (1997) estimate a luminosity of the order of 170-180 $L_\odot$ for component Aa and of 30-70 $L_\odot$ for component Ab.

Levreault (1988) detected a redshifted molecular outflow associated with V380 Ori. With the evidence that a jet is present there, a new higher angular resolution and
sensitivity molecular mapping of the region may provide valuable new information.

\subsection{The radio luminosity of V380 Ori}

The radio luminosity of radio jets, $S_\nu d^2$, where  $S_\nu$ is the flux density at 8 GHz in mJy and
$d$ is the distance to the source in kpc, is correlated with the bolometric luminosity of
the source, $L_{bol}$ (Anglada et al. 2015) by:

\begin{equation} {\biggl(\frac{S_\nu d^2}{\rm mJy~kpc^2}\biggr)} = 0.008 
\biggl(\frac{L_{\rm bol}}{L_\odot}\biggr)^{0.6}. \end{equation}

We first discuss the possibility that the radio jet is associated with 
component Aa of V380 Ori, that has a bolometric luminosity of $\sim$175 $L_\odot$ (Leinert et al. 1997).
Assuming that this source is located at a distance of 
460 pc (Reipurth et al. 2013), we expect a flux density of $\sim$0.84 mJy at $\sim$8 GHz from the correlation given above. 
This is a factor of 3.8 larger than the measured value of
0.22 mJy, indicating that this object is underluminous in the radio. This probably results from the fact that, with an age of about 2 million years (Alecian et al. 2009),
component Aa of V380 Ori is a more evolved object than those used to establish the correlation
(that typically have ages below a few times $10^5$ yr, Anglada et al. 2015).
This correlation most probably reflects the fact that in very young stars the luminosity is dominated by accretion, which in its turn is
correlated with the outflow 
activity that is traced by the radio free-free emission. In these very young objects the stellar contribution
to the bolometric luminosity is relatively unimportant. As the star evolves, the accretion decreases and the relative stellar contribution
becomes more important, making the star
radio-underluminous with respect to the Anglada et al. correlation. A similar underluminous radio jet has been recently found in AB Aur (Rodr\'\i guez et al. 2014).
In contrast, the nearby radio jet HH 1-2 VLA1
is a much younger class 0 protostar (age $< 10^5$ yr; Andr\'e et al. 2000) and has a flux density of $\sim$1.0 mJy at $\sim$8 GHz (Rodr\'\i guez et al. 2000).
With a bolometric luminosity of 23 $L_\odot$ (Fischer et al. 2010), this source is about 4 times overluminous in the radio with respect to
the correlation.

On the other hand, if we assume that the jet comes from component Ab (the infrared companion) that has a luminosity of $\sim$50 $L_\odot$,
the expected flux density would be $\sim$0.40 mJy at $\sim$8 GHz. This is only a factor of two larger than the measured value of 0.22 mJy.
However, in this case the Herbig Be star (component Aa) would be strongly underluminous in the radio, suggesting considerable evolution. In any case,
the data from which the correlation is derived shows considerable scatter and 
better and more abundant data are needed to clearly establish observationally if there is, as expected,
a correlation between radio luminosity and age for young stars. The search for this correlation is complicated, as in this case,
by the multiplicity of young stellar systems.

\section{Conclusions}

We present the analysis of archive VLA data as well as new high sensitivity Jansky VLA observations toward
V380 Ori. The main results of our study
can be summarized as follows.

1. We detect a radio counterpart to V380 Ori
that has spectral index, morphology, 
and lack of time variability all consistent with the source being a free-free jet.

2. The major axis of the radio source aligns, within the observational error, with the position angle of the proper 
motions of HH~222 and the direction from V380 Ori to the core
of HH 222. We propose that this radio jet traces the origin of the large scale HH outflow,
supporting the suggestion of Reipurth et al. (2013) that V380 Ori is the exciting source of the giant HH~222/HH~1041 system.

3. We cannot establish unambiguously if the jet originates from the Herbig Be star (component Aa) or from
its close (separation of  $0\rlap.{''}15$) infrared companion (component Ab). 
In any case, the radio luminosity of the jet is smaller than expected from the bolometric 
luminosity of either component.
We suggest that better and more abundant data are needed to clearly establish if there is
a decrease in thermal radio luminosity with age for young stars.

\acknowledgments

We thank an anonymous referee for valuable comments.
This research has made use of the SIMBAD database,
operated at CDS, Strasbourg, France.
LFR is grateful to CONACyT, Mexico and DGAPA, UNAM for their financial
support. GA acknowledges support from MINECO (Spain) grant AYA2014-57369-C3-3-P (co-funded with FEDER funds).

\clearpage

\end{document}